\documentclass[11pt,tightenlines]{revtex4}
\usepackage{amsmath}
\usepackage{amssymb}
\usepackage{bbm}
\usepackage{mathtools}
\usepackage{physics}
\usepackage{subfigure}
\usepackage{appendix}
\usepackage[utf8]{inputenc}
\usepackage[T1]{fontenc}

\def\be{\begin{equation}}
\def\ee{\end{equation}}

\def\bee{\begin{equation*}}
\def\eee{\end{equation*}}

\def\ba{\begin{equation}\begin{aligned}}
\def\ea{\end{aligned}\end{equation}}

\makeatletter
\newcommand{\pushright}[1]{\ifmeasuring@#1\else\omit\hfill$\displaystyle#1$\fi\ignorespaces}
\newcommand{\pushleft}[1]{\ifmeasuring@#1\else\omit$\displaystyle#1$\hfill\fi\ignorespaces}

\def\@thesubfigure{(\alph{subfigure})}
\makeatother

\usepackage[ddmmyyyy]{datetime}
\makeatletter
\def\Dated@name{}
\makeatother

\begin{document}

\title{Quantum critical point of the Ising chain from boundary effects}
\author{Oskar A. Pro\'s{}niak}
\affiliation{Jagiellonian University, Institute of Physics, {\L}ojasiewicza 11, 30-348 Krak\'ow, Poland}
\begin{abstract}
We propose two easy-to-study observables in the quantum Ising chain with open boundary conditions. They measure the length at which boundaries affect the longitudinal or transverse magnetization. We show that their finite-size scaling behaviour encodes the position of the quantum critical point and the universal scaling exponent $\nu$. The applicability of proposed observables in small systems is also discussed. We expect that our results will be useful in quantum simulation of spin systems.
\end{abstract}
\maketitle

\section{Introduction}
Quantum phase transitions (QPTs) 
are defined as the transitions between distinct phases of matter occurring at zero temperature at the thermodynamic limit \cite{Sachdev}. 
They result from the competition between basic interactions appearing in the system \cite{Continentino}. This makes them 
distinct from classical phase transitions, where the competition between internal energy and entropic contribution 
drives such transitions. 

Quantum critical phenomena can be observed in many different systems. It is natural to study them in solid state systems 
because  crystals are the physical realizations of lattices used in theoretical models
\cite{Sachdev,Coldea}. 
Solid state systems, however, are very complicated. 
Therefore, it is not always clear which theoretical model describes their behavior. 
In order to get rid of such ambiguities  and to study more exotic phases,
researchers have found ultracold atoms to be a promising setup \cite{Bloch, Greiner, Lewenstein, Monroe2}. 
Soon it has been realized that because of a great experimental ability to control atomic gases, these 
`artificial solids' could be used as quantum simulators \cite{Feynman, Bloch, Gross}.
By tuning their parameters, one can experimentally realize a wide range of theoretical models.

There is, however,  a price one has to pay for such controllability including 
tailoring  of interactions as well as  single-site preparation and measurement. Namely, 
currently realizable quantum simulators are composed of only  a few tens of  atoms or ions
\cite{Bernien, Monroe, Jurcevic, Flaschner}. 
As a result, they are not large enough to ensure that the thermodynamic limit from the definition of a QPT is well approximated. 
This implies that finite-size  predictions describing observables in such systems are much  needed \citep{Cardy}.

One of the parameters describing critical phenomena is the location of the critical point. 
The critical point 
is such a value of the
parameter of the Hamiltonian for which the ground state wave-function is non-analytic in the thermodynamically-large system.
It can be obtained in many different ways that differ in their sensitivity to
finite-size effects. For example, it can be extracted 
from the asymptotic behavior of the correlation functions 
(see \cite{Sachdev} for general introduction and \cite{Pfeuty, Lieb, Barouch} for Ising model results relevant for our work). 
However, due to the small size of quantum simulators, distant correlation functions 
are inaccessible in such systems. Moreover, the use of finite-size scaling procedure could be 
problematic as the system must be sufficiently large for its applicability.

Other approaches to finding the critical point, which are presumably less sensitive to the system size, 
include the fidelity approach (see \cite{Gu, Zanardi} for general introduction and \cite{Damski, DamskiRams} for Ising model results),  
quantum discord approach \cite{Sarandy}, and the studies of entanglement \cite{Osterloh}. 
A very unusual  method having  a broad spectrum of applicability was also proposed in \cite{Sen, Sen2}. 
It consists of applying Benford's law to physical observables.

\begin{figure}[t]
\includegraphics[scale=0.37]{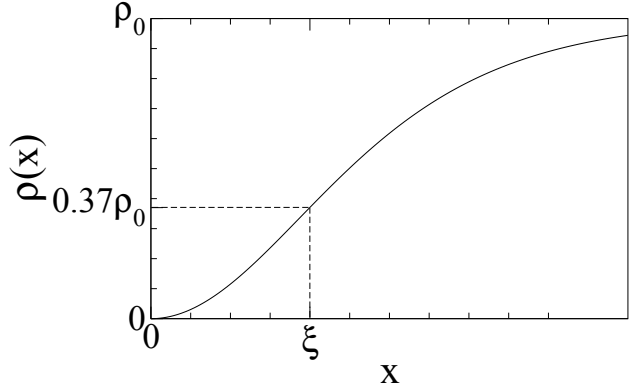}
\caption{Plot of the density of atoms in a Bose-Einstein condensate $\rho(x)=\rho_0\tanh^2\left[ x/(\sqrt{2}\xi) \right]$ illustrating the notion of the healing length $\xi$ at which $\rho(\xi)\approx 0.37\rho_0$. The condensate is subjected to boundary conditions: $\rho(0)=0$ and $\rho(\infty)=\rho_0$ \cite{Pethick}. It can be viewed as an atom density of the condensate trapped in a box potential near the boundary \cite{Pitaevski}.}\label{fig:healing_length}
\end{figure}

The objectives of this paper are twofold. 
First, we would like to propose an observable, which should be easy-to-study in quantum simulators, 
whose scaling properties encode the critical value of the parameter driving the transition. 
Second, we would like to provide its detailed discussion in the open quantum Ising chain. The proposed observable measures the length of the boundary effect on the magnetization. Such an observable is analogues to the healing length of the Bose-Einstein condensate  (Fig. \ref{fig:healing_length}), which describes the distance over which a localized perturbation (like an infinite potential barrier) changes the density of atoms \cite{Pethick}.
 
\section{Finite-size effects}
Quantum phase transitions are traditionally considered in a thermodynamic limit. Therefore, it is difficult to measure the critical value of the parameter driving the transition in a small system. In such a case, one can extrapolate it using finite-size scaling theory \cite{Cardy}.

Let us consider an infinitely large system where quantum phase transition occurs at the driving parameter $g=g_c$. There exists a characteristic length scale $\xi$ such as a correlation length. We assume that in the regime of $g$ close to the critical value
\be
\xi \propto \abs{g - g_c}^{-\nu}, \label{eq:xi}
\ee
where $\nu$ is a universal critical exponent.
Suppose now there is an observable $O_\infty$ which in this regime diverges algebraically with some critical exponent $\gamma$. Therefore
\be
O_\infty \propto \xi^{\gamma / \nu}. \label{eq:scaling_ans}
\ee
If we would now consider a finite version of that system of the length $L$, this proportionality must be altered by the scaling function $\Phi$ such that finite-system version of $O_\infty$
\be
O_L \propto \xi^{\gamma / \nu} \Phi (\xi/L), \label{eq:finite_obs}
\ee
which is the essence of the finite-size scaling hypothesis. Making use of \eqref{eq:xi} and introducing another scaling function $\tilde{\Phi}$ one can rewrite \eqref{eq:finite_obs} as
\be
O_L \propto L^{\gamma/\nu} \tilde{\Phi}(|g-g_c|^{-\nu}/L). \label{eq:obs_scaling}
\ee
In a case of an infinite system, $O_\infty$ would be non-analytical at $g=g_c$. In a finite system, however, $O_L$ has an extremum near the critical point at $g=g^*$. We therefore conclude that
\be
|g^*-g_c|=x_0^{-1/\nu} L^{-1/\nu} \propto L^{-1/\nu} \propto N^{-1/\nu}, \label{eq:scaling}
\ee 
where $x_0$ is a point where $\tilde{\Phi}$ takes its extremal value and $N=L/a$ with $a$ being the lattice constant.

One has to be a bit more careful if the observable of interest is logarithmically divergent, i.e.
\be
O_\infty = \alpha \ln(\xi),
\ee
where $\alpha$ is a constant of proportionality.

Let us now consider an observable $\tilde{O}_\infty$ which is an exponentiation of $O_\infty$. In a finite system we have
\be
\tilde{O}_L = \tilde{O}_\infty \Phi (\xi/L) = \exp(O_\infty) \Phi (\xi/L),
\ee
where $\Phi$  is an appropriate scaling function. Therefore, introducing another scaling function $\tilde{\Phi}$
\begin{multline}
O_L = \ln(\tilde{O}_L) = \alpha \ln(\xi) + \ln[\Phi(\xi/L)] = \alpha \ln(L \cdot \xi /L) + \ln[\Phi(\xi/L)] =\\
 = \alpha \ln(L) +\ln[\left( \xi/L \right)^\alpha \Phi(\xi/L)] =  \alpha \ln(L) + \tilde{\Phi}(\xi/L).
\end{multline}
Using the same reasoning as before, we again obtain \eqref{eq:scaling}.

\section{Ising model}
The model of interest is the one-dimensional quantum Ising chain in the external transverse magnetic field with open boundary conditions. Its Hamiltonian reads \citep{Sachdev}
\be
H = -\sum_{i=1}^{N-1} \sigma_i^x\sigma_{i+1}^x-g\sum_{i=1}^N \sigma_i^z, \label{eq:H}
\ee
where $\sigma_i^{x,z}$ are Pauli matrices describing the $i$th spin-$1/2$, $N$ is the number of spins, and $g$ is a parameter describing the strength of the external field.

In a thermodynamic limit, there exists a quantum phase transition driven by the external field. For $0\leq g < 1$ the system is in the ferromagnetic phase, while for $g>1$ it is in the paramagnetic phase.

The model has many physical realizations in different solid state setups \cite{Dutta}. Recently, cold ions systems have grasped the attention of experimentalists \cite{Porras}. The version of Ising model with long-range interactions has been successfully implemented in linear ion chains \cite{Monroe2} with at most 53 effective spins-1/2 \cite{Monroe}, which makes our discussion experimentally relevant.

Following \cite{Young} Hamiltonian \eqref{eq:H} can be rewritten in the useful second-quantized form
\be
H=\Psi^\dag \tilde{H} \Psi, \label{eq:sqHamiltonian}
\ee
where
\be
\tilde{H}=\mqty(A & B \\ -B & -A),
\ee
with $A_{i\:i}=g,\,A_{i\:i+1}=A_{i+1\:i}=B_{i\:i+1}=-B_{i+1\:i}=-1/2 $ (where we have corrected sign error in $B_{i\:i+1}$ and $B_{i+1\:i}$)
and
\be
\Psi_i=
\begin{cases}
c_i & 1\leq i \leq N\\
c_i^\dag & N < i \leq 2N
\end{cases} .
\ee
Introduced creation and annihilation operators come from the Jordan-Wigner transformation
\be
\sigma_i^z=1-2c_i^\dag c_i,\quad \sigma_i^x=\left(c_i+c_i^\dag\right) \prod_{j<i} \left(1-2c_j^\dag c_j\right). \label{eq:J-W}
\ee

\begin{figure}[t]
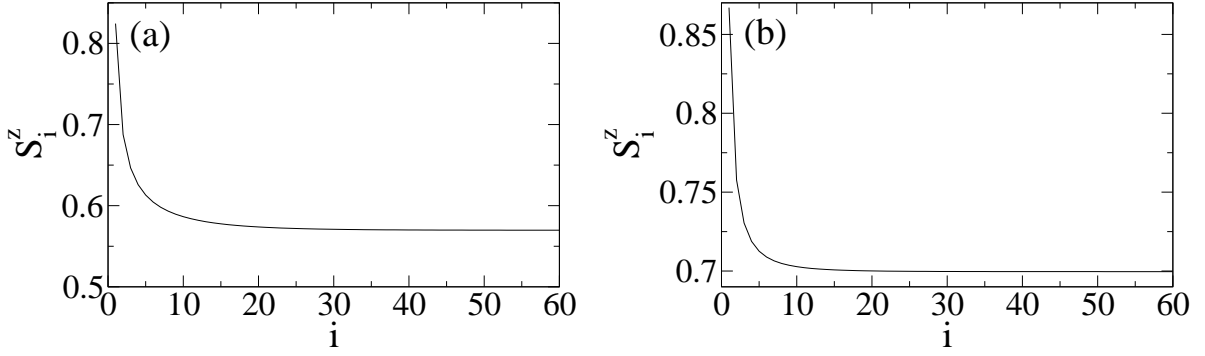

\subfigure{
\includegraphics[scale=0.35]{fig2.eps}\label{fig:sz_9500}
}
\subfigure{
\includegraphics[scale=0.35]{fig3.eps}\label{fig:sz_10500}
}
\caption{Magnetization $S_i^z$ obtained in numerical computation for the system of length $N=10^3$. The index $i$ enumerates lattice sites. The boundary is at $i=1$ and $i=N$. Panel (a): magnetization for $g=0.95$. Panel (b): magnetization for $g=1.05$. Data points are connected by lines.}\label{fig:sz}
\end{figure}

Hamiltonian \eqref{eq:sqHamiltonian} can be easily diagonalized such that (using Einstein summation convention)
\be
H=\Psi_i^\dag \tilde{H}_{ij} \Psi_j=\Psi_i^\dag \beta_{ik}D_{kl}\beta_{lj}^T\Psi_j=\Psi_i^\dag \beta_{ik}D_{kk}\beta_{kj}^T\Psi_j=
\Phi_k^\dag D_{kk} \Phi_k = \tilde{\epsilon}_k \Phi_k^\dag\Phi_k, \label{eq:diagHamiltonian}
\ee
where $\beta$ is an orthogonal transition matrix, $\Phi = \beta^T \Psi$ and $D = \text{diag}(\{\tilde{\epsilon}_k\})$ with \cite{YanHe}
\be
\tilde{\epsilon}_k=\pm \sqrt{\left[g-\cos(\theta_k)\right]^2+\sin^2(\theta_k)},
\ee
where $\theta_k$ are real roots of the equation for $\theta$
\be
g\sin\left((N+1)\theta\right)=-\sin(N\theta),\quad 0<\theta<\pi.
\ee
The set $\{\tilde{\epsilon}_k\}$ is invariant under the change of sign of all elements as it comes from the symmetries of $\tilde{H}$. Indeed, if
\be
\psi_k = \mqty(u_k \\ v_k)
\ee
is a normalized eigenvector to an eigenvalue $\epsilon_k \geq 0$, then  $\mqty(v_k \\ u_k)$ is a normalized eigenvector to an eigenvalue $-\epsilon_k$. We can therefore choose
\be
\beta = \mqty(u_1 & \dots & u_N & v_1 & \dots & v_N \\ v_1 & \dots & v_N & u_1 & \dots & u_N),
\ee
and assume that eigenvalues $\epsilon_k$ are in an ascending order. Moreover, $\beta$ can always be chosen to be real. It is then natural to define Bogoliubov operators
\be
\gamma_k = \Phi_k = \sum_{j=1}^{2N} \beta_{kj}^T \Psi_j = \sum_{i=1}^N \left( u_{ki}c_i + v_{ki}c_i^\dag \right) \quad \text{for }1\leq k \leq N,
\ee
where $u_{ki}$ and $v_{ki}$ are the $i$-th components of the $u_k$ and $v_k$ vectors, respectively. Thanks to the arrangement of matrix $\beta$ it easily follows that $\gamma_k^\dagger = \Phi_{k+N}$.
Anticommutation relations are fulfilled due to the properties of the diagonalized matrix and the normalization of its eigenvectors. In the end, we arrive at
\be
H = \sum_{k=1}^N \epsilon_k\left(2\gamma_k^\dag\gamma_k-1\right). \label{eq:finalHamiltonian}
\ee

\section{Magnetization}
The transverse magnetization in the ground state (Fig. \ref{fig:sz}) reads 
\begin{multline}
S_i^z = \expval{\sigma_i^z}_{GS}=\expval{1-2c_i^\dag c_i}_{GS}=1-2\expval{\Psi_i^\dag \Psi_i}_{GS}=1-2\sum_{j,k=1}^{2N} \expval{\left( \beta_{ik}\Phi_k \right)^\dag \beta_{ij}\Phi_j}_{GS}=\\
=1-2\sum_{j=N+1}^{2N}(\beta_{ij})^2, \label{eq:tr_mag}
\end{multline}
where $\expval{\dots}_{GS}$ is the ground state expectation value.

\begin{figure}[t]
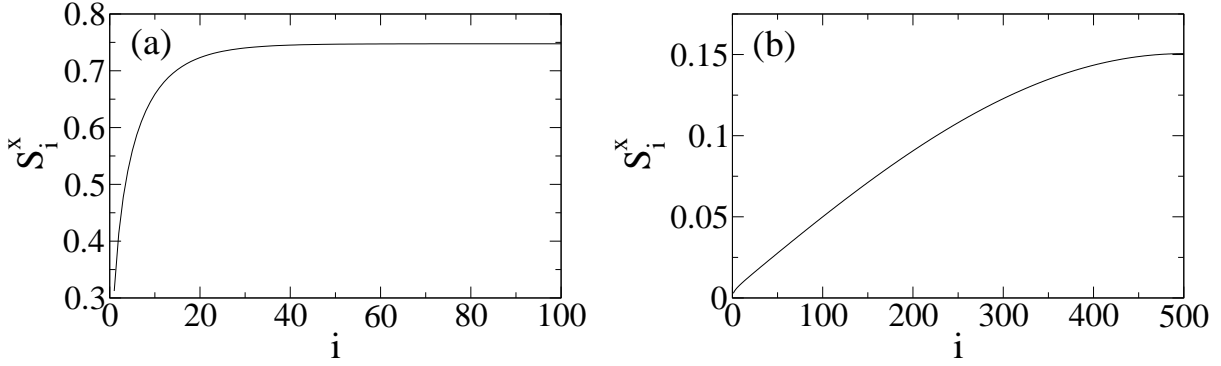

\subfigure{
\includegraphics[scale=0.35]{fig4.eps}\label{fig:sx_9500}
}
\subfigure{
\includegraphics[scale=0.35]{fig5.eps}\label{fig:sx_10500}
}
\caption{Magnetization $S_i^x$ obtained in numerical computation for the system of length $N=10^3$. The index $i$ enumerates lattice sites. The boundary is at $i=1$ and $i=N$. Panel (a): magnetization for $g=0.95$. Panel (b): magnetization for $g=1.05$. Data points are connected by lines. Note the different range of the horizontal axis in each panel.}\label{fig:sx}
\end{figure}

The longitudinal magnetization in any eigenstate of \eqref{eq:finalHamiltonian} is equal to zero by Wick's theorem, because operators $c_i$ and $c_i^\dag$, which can be linearly mapped onto $\gamma_i$ and $\gamma_i^\dag$, appear an odd number of times in $\sigma_i^x$. However, one can consider adding an infinitesimally small symmetry-breaking field along the spin interactions. Such a field will mix two lowest eigenstates such that the resulting superposition can be written as \cite{Platini}
\be
\ket{\psi} = \frac{\ket{GS} + \alpha e^{i \phi} \gamma_1^\dag \ket{GS}}{\sqrt{1+\alpha^2}}, \quad \alpha > 0, \quad \phi \in [0,2\pi).
\ee
The longitudinal magnetization in this state reads
\be
\expval{\sigma_i^x}_{\psi} = \frac{2\alpha \cos{\phi}}{1+\alpha^2}\expval{\gamma_1 \sigma_i^x}_{GS}.
\ee
It has its maximum at $\alpha = 1$ and $\phi = 0$, and we can treat the following expression as the longitudinal magnetization of the perturbed system (Fig. \ref{fig:sx})
\be
S_i^x = \expval{\gamma_1 \sigma_i^x}_{GS}. \label{eq:ln_mag}
\ee
Using the Jordan-Wigner transformation \eqref{eq:J-W} and defining
\be
a_i = c_i+c_i^\dag,\,b_i = i(c_i^\dag - c_i), \,\{a_i,b_i\}=0,
\ee
allows us to obtain
\be
|S_i^x| = |\expval{\gamma_1a_ib_{i-1}a_{i-1}\dots b_1a_1}_{GS}|,
\ee
which can be then computed with the use of Wick's theorem \cite{Platini}, because operators $a_i$ and $b_i$ are linear combinations of operators $\gamma_i$ and $\gamma_i^\dag$, in which the Hamiltonian \eqref{eq:finalHamiltonian} is diagonal. Conveniently, it can be done with the use of the identity
\be
\expval{\alpha_1 \alpha_2 \dots \alpha_M} = \text{Pf} \mqty(0 & \expval{\alpha_1 \alpha_2} & \expval{\alpha_1 \alpha_3} & \expval{\alpha_1 \alpha_4} & \dots & \expval{\alpha_1 \alpha_M}  \\ & 0 & \expval{\alpha_2 \alpha_3} & \expval{\alpha_2 \alpha_4} & \dots & \expval{\alpha_2 \alpha_M} \\ & & 0& \expval{\alpha_3 \alpha_4} & \dots & \expval{\alpha_3 \alpha_M} \\ & & & & & \vdots \\ & & & & & 0),
\ee
where the used matrix is skew-symmetric and $\alpha_i$ are creation or annihilation operators \cite{Pfaffian}.

\begin{figure}[t]
\subfigure{
\includegraphics[scale=0.35]{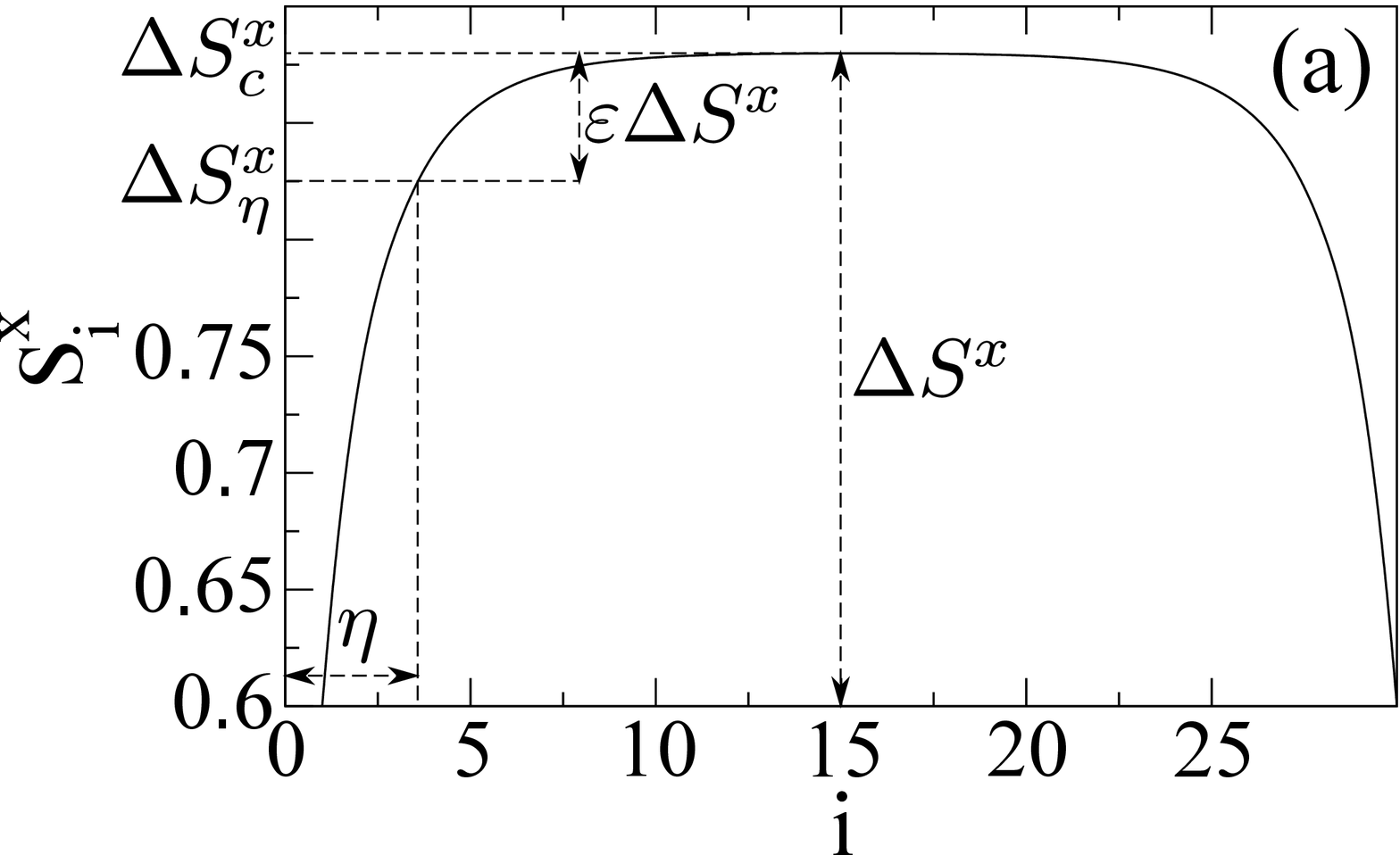}
}
\subfigure{
\includegraphics[scale=0.35]{fig7.eps}
}
\caption{Panel (a): illustration explaining quantities used in definition of the observable $\eta$. The system has a length of $N=30$. Arbitrary value of $\varepsilon$ has been used. Panel (b): definition of the observable $\eta$. The system has a length of $N=10^3$. To illustrate the method, we use here $\varepsilon=10\%$.}\label{fig:obs}
\end{figure}

\section{The observable}
We propose the observable $\eta$ similar to the healing length (Fig. \ref{fig:healing_length}) but in the finite Ising chain. It is not an observable in the strict sense, but we will stick to this term in further description. It is the length, measured from the end of the chain in lattice units, at which the longitudinal magnetization $S_i^x$ differs form its bulk value $S_c^x$ by a tiny percent $\varepsilon$ of the magnetization range $\Delta S^x$. $\Delta S^x$ is defined as the difference between $S^x$ magnetization at the end and at the bulk. It is illustrated in Fig.~\ref{fig:obs}.

\begin{figure}[t]
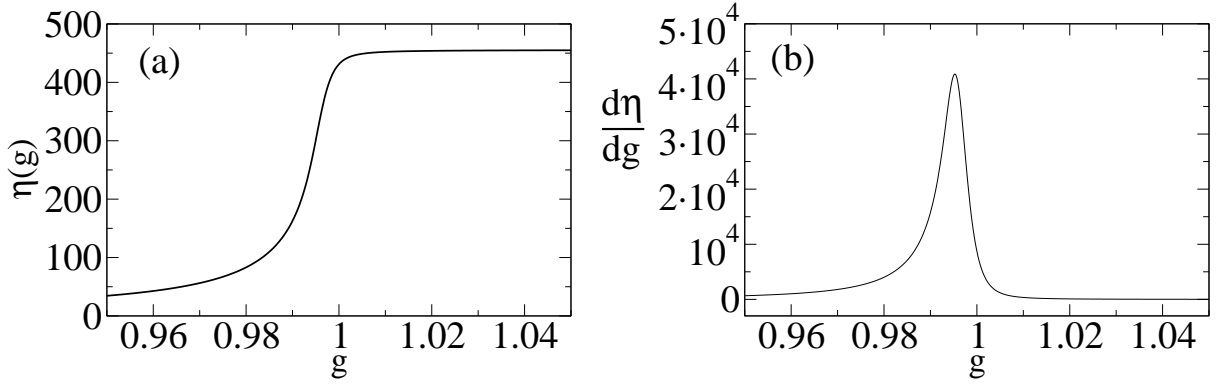

\subfigure{
\includegraphics[scale=0.36]{fig8.eps}\label{fig:corr_sx_1000}
}
\subfigure{
\includegraphics[scale=0.36]{fig9.eps}\label{fig:dcorr_sx_1000}
}
\caption{Results for the proposed observable for $S^x$ magnetization for the system of length $N=10^3$ and $\varepsilon=1\%$. Panel~(a): $\eta(g)$. Panel~(b): first derivative of $\eta(g)$ with clearly visible maximum. Data points are connected by lines.}
\end{figure}

\begin{figure}[t]
\includegraphics[scale=0.37,clip]{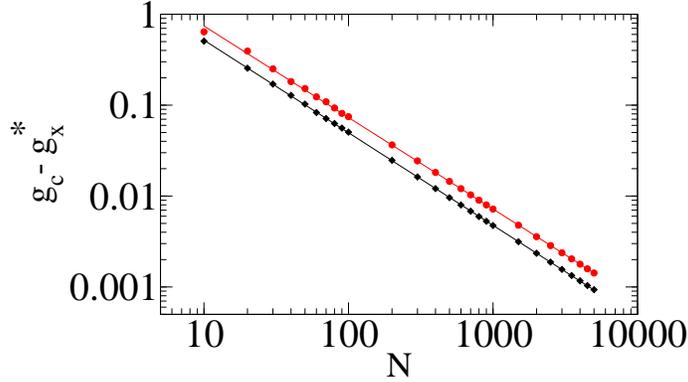}
\caption{Logarithmic plot of the distance of the maximum of $\dv{g}\eta(g)$ from the critical point $g_c-g_x^*$ as a function of $N$. The black line is a linear fit $\ln(g_c-g_x^*)=1.77(2)-1.030(3)\ln(N)$ to data points for $\varepsilon=1\%$, while black diamonds are exemplary data points. Similarly, the red line is a linear fit $\ln(g_c-g_x^*)=2.02(2)-1.006(2)\ln(N)$ to data points for $\varepsilon=0.1\%$, while red dots are exemplary data points. We used over 40 points for each fit. The fitted coefficients and their standard errors, both here and in other plots in this work, come from LinearModelFit function from \cite{Mathematica}.}\label{fig:g*}
\end{figure}

Formally, $\eta$ for an Ising chain of length $N$ in the external magnetic field of strength $g$ is defined as
\be
\frac{S_c^x - S_\eta^x}{\Delta S^x} = \varepsilon,\label{eq:eta}
\ee
where $S_c^x = S_{\lfloor N/2 \rfloor}^x$, $\Delta S^x = S_{\lfloor N/2 \rfloor}^x - S_{1}^x$, and $S_\eta^x$ interpolates the system magnetization at non-necessarily integer distance $\eta$ from the boundary (see Appendix for details). The function ${\lfloor x \rfloor}$ returns the greatest integer smaller than its argument $x$.

Treating $\eta$ as a function of $g$ (Fig. \ref{fig:corr_sx_1000}), it was numerically shown that its first derivative $\dv{g}\eta(g)$ (Fig. \ref{fig:dcorr_sx_1000}) has a maximum at $g_x^*$, which obeys scaling relation \eqref{eq:scaling}. From the fit (Fig.~\ref{fig:g*}) we see that $|g_x^*-g_c|\propto N^{-1.030(3)}$ for $\varepsilon = 1\%$ and $|g_x^*-g_c|\propto N^{-1.006(2)}$ for $\varepsilon = 0.1\%$. Obtained values of the critical exponent read $\nu = 0.971(3) \approx 1$ for $\varepsilon = 1\%$ and $\nu = 0.994(2)\approx 1$ for $\varepsilon = 0.1\%$, which is in good agreement with the theory \cite{Sachdev,Barouch,nu} and suggests asymptotic in $\varepsilon \rightarrow 0$ convergence to $\nu = 1$.

\begin{figure}[t]
\includegraphics[scale=0.37]{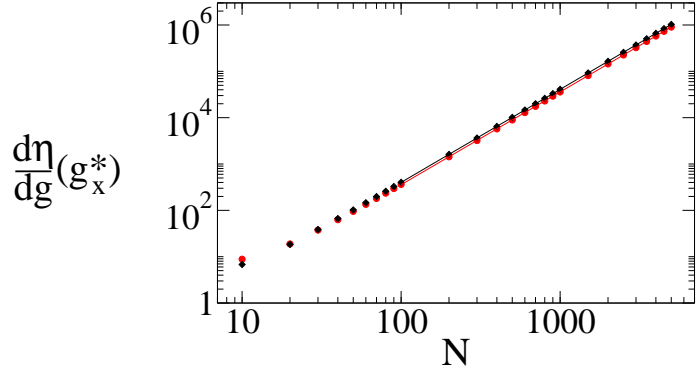}
\caption{Logarithmic plot of the distance of the maximum of $\dv{g}\eta(g_x^*)$ as a function of $N$. The black line is a linear fit $\ln[\dv{g}\eta(g_x^*)]=-3.246(2)+2.0070(3)\ln(N)$ to data points for $\varepsilon=1\%$, while black diamonds are exemplary data points. Similarly, the red line is a linear fit $\ln[\dv{g}\eta(g_x^*)]=-3.339(4)+2.0021(6)\ln(N)$ to data points for $\varepsilon=0.1\%$, while red dots are exemplary data points. We used over 30 points in between $N=100$ and $N=5000$ for each fit.}\label{fig:detaMax}
\end{figure}

We have also checked the scaling behaviour of the $\dv{g}\eta(g_x^*)$. According to \eqref{eq:obs_scaling} and \eqref{eq:scaling}, the value of the observable at $g_x^*$ obey power law scaling with $N$ with the characteristic exponent $\gamma/\nu$. We have made linear fits to data points in the area of linearity (see Fig.~\ref{fig:detaMax} for details) obtaining $\dv{g}\eta(g_x^*) \propto N^{2.0070(3)}$ for $\varepsilon=1\%$ and $\dv{g}\eta(g_x^*) \propto N^{2.0021(6)}$ for $\varepsilon=0.1\%$. It suggests that $\gamma / \nu \approx 2$. This agrees with intuition, since $\eta$ measures a length and the unit of $g$ is the inverse of the unit of length, therefore $\dv{g}\eta(g_x^*)$ is expressed in terms of length squared. In the vicinity of critical point, we assume no length scale other than $N$ \cite{Cardy}, therefore $\dv{g}\eta(g_x^*)$ must be proportional to $N^2$.

\begin{figure}[t]
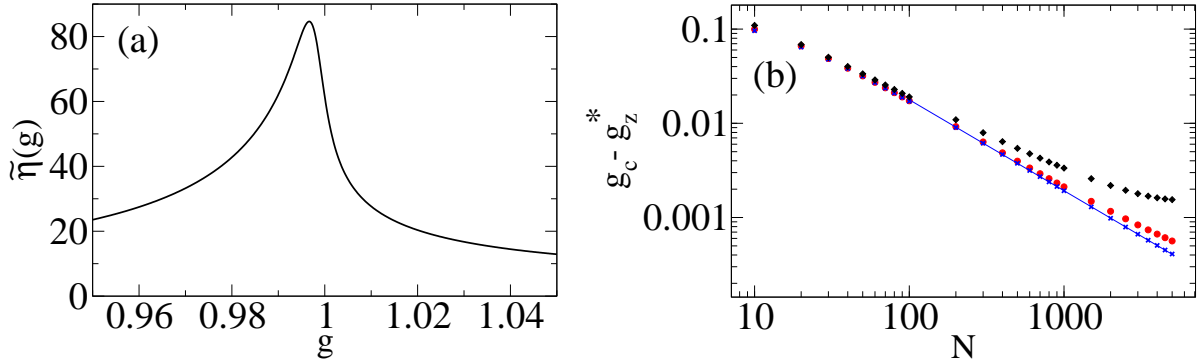

\subfigure{
\includegraphics[scale=0.36,clip]{fig12.eps}\label{fig:corr_sz_1000}
}
\subfigure{
\includegraphics[scale=0.36,clip]{fig13.eps}\label{fig:sz_fit}
}
\caption{Results for proposed observable for $S^z$ magnetization for the system of length $N=10^3$ and $\varepsilon=1\%$. Panel~(a): $\tilde{\eta}(g)$. Data points are connected by lines. Panel~(b): logarithmic plot of the distance of the maximum of $\tilde{\eta}(g)$ from the critical point $g_c-g_z^*$ as a function of $N$. Black diamonds are exemplary data points for $\varepsilon=1\%$. Red dots are exemplary data points for $\varepsilon=0.1\%$. The blue line is a linear fit $\ln(g_c-g_z^*)=0.417(2)-0.9656(3)\ln(N)$ to data points for $\varepsilon=0.01\%$, while blue crosses are exemplary data points. Data points for smaller values of $\varepsilon$ are not included since they practically overlap with points for $\varepsilon=0.01\%$.}
\end{figure}

Similar simulations have been made with the observable $\tilde{\eta}$ defined as follows
\be
\frac{S_c^z - S_{\tilde{\eta}}^z}{\Delta S^z} = \varepsilon,
\ee
where $S_c^z = S_{\lfloor N/2 \rfloor}^z$, $\Delta S^z = S_{\lfloor N/2 \rfloor}^z - S_{1}^z$, and $S_{\tilde{\eta}}^z$ interpolates the system magnetization at non-necessarily integer distance $\tilde{\eta}$ from the boundary.

Quite interestingly, with such a definition of the proposed observable no differentiation is needed in order to obtain peaked function and therefore $g_z^*$, as can be seen in Fig.~\ref{fig:corr_sz_1000}. To check scaling relation \eqref{eq:scaling}, we have plotted obtained values of $g_z^*$ (Fig.~\ref{fig:sz_fit}). Differently than it was in the case of $g_x^*$ (Fig.~\ref{fig:g*}), not all data points lay on a straight line. We have made simulations for several values of $\varepsilon$. For $\varepsilon\leq 0.01\%$ there exist an area of linearity for $500\leq N\leq5000$. We have made linear fits to over 90 data points lying in this area obtaining $|g_x^*-g_c|\propto N^{-0.9656(3)}$ for $\varepsilon = 0.01\%$, $|g_x^*-g_c|\propto N^{-0.9848(2)}$ for $\varepsilon = 0.001\%$ and $|g_x^*-g_c|\propto N^{-0.9869(3)}$ for $\varepsilon = 0.0001\%$. These lead to values of $\nu$ collected in Tab.~\ref{tab:all}.

\begin{figure}[t]
\includegraphics[scale=0.37]{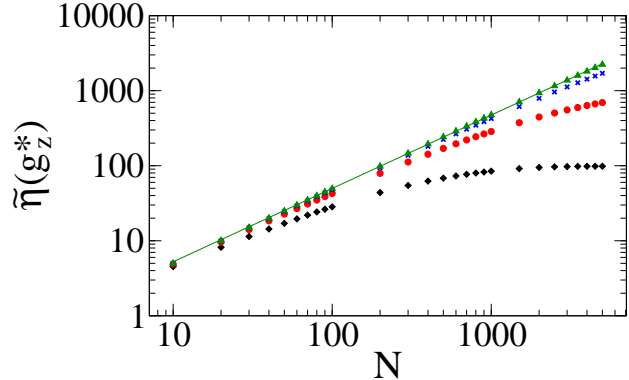}
\caption{Logarithmic plot of the distance of the maximum of $\tilde{\eta}(g_z^*)$ as a function of $N$. Black diamonds, red dots, blue crosses, and green triangles are exemplary data points for $\varepsilon=1\%$, $\varepsilon=0.1\%$, $\varepsilon=0.01\%$, and $\varepsilon=0.001\%$ respectively. The green line is a linear fit $\ln\left( \tilde{\eta}(g_z^*)\right) = -0.594(4)+0.9777(7)\ln(N)$ to data points for $\varepsilon=0.001\%$. Data points for $\varepsilon = 0.0001\%$ are not included since they practically overlap with points for $\varepsilon=0.001\%$.}\label{fig:etaMax}
\end{figure}

We have also checked the behaviour of $\tilde{\eta}(g_z^*)$. Since $\tilde{\eta}$ is measuring a length, it is expected to be proportional to $N$ in the vicinity of the critical point, i.e. $\gamma / \nu =1$. This is indeed the case, as can be seen from Fig.~\ref{fig:etaMax}, for small values of $\varepsilon$. We have made linear fits to these data points ($140$ points between $N=10$ and $N=5000$) obtaining $\tilde{\eta}(g_z^*) \propto N^{0.9777(7)}$ for $\varepsilon = 0.001\%$ and $\tilde{\eta}(g_z^*) \propto N^{0.9929(2)}$ for $\varepsilon = 0.0001\%$. For $\varepsilon\geq 0.01\%$ data points are not lying on a straight line, unlike in the case of $\dv{g}\eta(g_x^*)$.

\begin{table}[!ht]
\caption{Approximate value of $\nu$ obtained from scaling ansatz applied to $g_x^*$ and $g_z^*$.} \label{tab:all}
\begin{tabular}{c | c c}
$\varepsilon$ & $\nu[g_x^*]$ & $\nu[g_z^*]$ \\ 
\hline
$1\%$ & 0.971(3) & - \\
$0.1\%$ & 0.994(2) & - \\
$0.01\%$ & - & 1.0357(3) \\
$0.001\%$ & - & 1.0154(2) \\
$0.0001\%$ & - & 1.0133(3)
\end{tabular}
\end{table}

\begin{table}[!ht]
\caption{Differences $g_c-g_x^*$ and $g_c-g_z^*$ for small systems and $\varepsilon=1\%$.} \label{tab:small}
\begin{tabular}{c | c c}
$N$ & $g_c-g_x^*$ & $g_c-g_z^*$ \\ 
\hline
10&0.6407&0.1095 \\
20&0.2580&0.06851 \\
30&0.1768&0.05026 \\
40&0.1336&0.04006 \\
50&0.1070&0.03350 
\end{tabular}
\end{table}

These numerical simulations show that both observables are well suited for searching for the critical point location. They also illustrate that it is possible to obtain critical exponent $\nu$ from their finite-size scaling. Moreover, they are sensitive to QPT even in small systems. In Tab.~\ref{tab:small} we also collected differences $g_c-g_x^*$ and $g_c-g_z^*$ for a few $N \leq 50$. For $N=50$ and $\varepsilon=1\%$ the critical point location is approximated with $10\%$ and $3\%$ accuracy without extrapolation by $g_x^*$ and $g_z^*$, respectively. 

\section{Summary}
In this work, we have proposed and studied two observables measuring a length at which boundaries affect the magnetization in the finite-size Ising model. They exhibit extrema which can be used for finding the position of the critical point. 

The first observable is defined for the longitudinal magnetization. We have shown that the position of the extremum of its first derivative scales with the size of the system according to the universal power law. Moreover, value of this extremum also scales with the system size as it is expected from dimensional analysis in the vicinity of the critical point.

The second observable is defined for the transverse magnetization. Numerical studies of its scaling behaviour show that this observable is intrinsically different from the one defined for longitudinal magnetization. It itself has an extremum which can be used for extrapolating the position of the critical point. However, its location is only approximately governed by the power law scaling. Similarly, the value of this extremum scales with the system size as it is expected from dimensional analysis only approximately.

We believe that proposed observables may be a computationally efficient and therefore a very useful tool for locating critical points in spin systems while at the same time providing an intuitive description of the finite-size effects.

\begin{center}
{\bf ACKNOWLEDGMENTS}
\end{center}

I would like to thank Bogdan Damski for inspiration, encouragement and his help with writing this paper and Mateusz {\L}\k{a}cki for his help with numerical simulations. This work was supported by the Polish National Science Centre (NCN) grant DEC\nobreakdash--2016/23/B/ST3/01152.

\appendix
\renewcommand{\thesection}{}
\section{}
In the numerical computation magnetization \eqref{eq:tr_mag} and \eqref{eq:ln_mag} was obtained following the described way of diagonalizing the Hamiltonian \eqref{eq:H} (Fig.~\ref{fig:sz}, Fig.~\ref{fig:sx}).

In order to calculate $\eta$ (Fig.~\ref{fig:corr_sx_1000}), we transformed equation \eqref{eq:eta} into
\be
S_c^x-S_{\eta}^x-\varepsilon \Delta S^x = 0.
\ee
The left-hand-side defines a function whose zero has to be found. We introduced the Piecewise Cubic Hermite Interpolating Polynomial \cite{python} for pairs
\be
\left\{ \left( i, S_c^x -S_i^x - \varepsilon \Delta S^x  \right) \right\}.
\ee
Then, using Brent's method \cite{python} its root corresponding to $\eta$ was found. It was done for different values of $L$, $g$ and $\varepsilon$.

To get $g_x^*$, the symmetric numerical derivative $\dv{g}\eta(g)$ was calculated (Fig.~\ref{fig:dcorr_sx_1000}). The vertex of a parabola constructed through three points from the set $\left\{\left( g, \dv{g}\eta(g) \right)\right\}$ with the largest value of $\dv{g}\eta(g)$ was assigned to value of $g_x^*$.

We proceeded with $\tilde{\eta}(g)$ as previously with the first derivative of $\eta(g)$.


\begin{thebibliography}{36}
\expandafter\ifx\csname natexlab\endcsname\relax\def\natexlab#1{#1}\fi
\expandafter\ifx\csname bibnamefont\endcsname\relax
  \def\bibnamefont#1{#1}\fi
\expandafter\ifx\csname bibfnamefont\endcsname\relax
  \def\bibfnamefont#1{#1}\fi
\expandafter\ifx\csname citenamefont\endcsname\relax
  \def\citenamefont#1{#1}\fi
\expandafter\ifx\csname url\endcsname\relax
  \def\url#1{\texttt{#1}}\fi
\expandafter\ifx\csname urlprefix\endcsname\relax\def\urlprefix{URL }\fi
\providecommand{\bibinfo}[2]{#2}
\providecommand{\eprint}[2][]{\url{#2}}

\bibitem[{Sac()}]{Sachdev}
\bibinfo{note}{S. Sachdev, {\it Quantum Phase Transitions} (Cambridge
  University Press, 2011).}

\bibitem[{Con()}]{Continentino}
\bibinfo{note}{M. A. Continentino, {\it Quantum Scaling in Many-Body Systems}
  (World Scientific Publishing, 2001).}

\bibitem[{Col()}]{Coldea}
\bibinfo{note}{R. Coldea {\it et al.}, Science {\bf 327}, 177 (2010).}

\bibitem[{Blo()}]{Bloch}
\bibinfo{note}{I. Bloch {\it et al.}, Nature Phys. {\bf 8}, 267 (2012)}.

\bibitem[{Gre()}]{Greiner}
\bibinfo{note}{M. Greiner {\it et al.}, Nature {\bf 415}, 39 (2002).}

\bibitem[{Lew()}]{Lewenstein}
\bibinfo{note}{M. Lewenstein {\it et al.}, Adv. Phys. {\bf 56}, 243 (2007)}.

\bibitem[{Mon({\natexlab{a}})}]{Monroe2}
\bibinfo{note}{S. Korenblit {\it et al.}, New J. Phys. {\bf 14}, 095024
  (2012)}.

\bibitem[{Fey()}]{Feynman}
\bibinfo{note}{R. P. Feynman, Int. J. Th. Phys. {\bf 21}, 467 (1982)}.

\bibitem[{Gro()}]{Gross}
\bibinfo{note}{C. Gross and I. Bloch, Science {\bf 357}, 995 (2017)}.

\bibitem[{Ber()}]{Bernien}
\bibinfo{note}{H. Bernien {\it et al.}, Nature {\bf 551}, 579 (2017).}

\bibitem[{Mon({\natexlab{b}})}]{Monroe}
\bibinfo{note}{J. Zhang {\it et al.}, Nature {\bf 551}, 601 (2017).}

\bibitem[{Jur()}]{Jurcevic}
\bibinfo{note}{P. Jurcevic {\it et al.}, Phys. Rev. Lett. {\bf 119}, 080501
  (2017)}.

\bibitem[{Fla()}]{Flaschner}
\bibinfo{note}{N. Fl\"aschner {\it et al.}, Nature Phys. {\bf 14}, 265 (2018)}.

\bibitem[{Car()}]{Cardy}
\bibinfo{note}{J. Cardy, {\it Finite-Size Scaling} (North-Holland, 1988).}

\bibitem[{Pfe()}]{Pfeuty}
\bibinfo{note}{P. Pfeuty, Ann. Phys. {\bf 57}, 79 (1970)}.

\bibitem[{Lie()}]{Lieb}
\bibinfo{note}{E. Lieb {\it et al.}, Ann. Phys. {\bf 16}, 407 (1961)}.

\bibitem[{Bar()}]{Barouch}
\bibinfo{note}{E. Barouch and B. M. McCoy, Phys. Rev. A {\bf 3}, 2 (1971)}.

\bibitem[{Gu()}]{Gu}
\bibinfo{note}{S.-J. Gu, Int. J. Mod. Phys. B {\bf 24}, 4371 (2010)}.

\bibitem[{Zan()}]{Zanardi}
\bibinfo{note}{P. Zanardi and N. Paunkovi\'c, Phys. Rev. E {\bf 74}, 031123
  (2006)}.

\bibitem[{Dam({\natexlab{a}})}]{Damski}
\bibinfo{note}{B. Damski, Phys. Rev. E {\bf 87}, 052131 (2013)}.

\bibitem[{Dam({\natexlab{b}})}]{DamskiRams}
\bibinfo{note}{M. M. Rams and B. Damski, Phys. Rev. Lett. {\bf 106}, 055701
  (2011)}.

\bibitem[{Sar()}]{Sarandy}
\bibinfo{note}{M. S. Sarandy, Phys. Rev. A {\bf 80}, 022108 (2009)}.

\bibitem[{Ost()}]{Osterloh}
\bibinfo{note}{A. Osterloh {\it et al.}, Nature {\bf 416}, 608 (2002)}.

\bibitem[{Sen({\natexlab{a}})}]{Sen}
\bibinfo{note}{A. Sen and U. Sen, Europhys. Lett. {\bf 95}, 50008 (2011)}.

\bibitem[{Sen({\natexlab{b}})}]{Sen2}
\bibinfo{note}{A. D. Rane {\it et al.}, Phys. Rev. E {\bf 90}, 022144 (2014)}.

\bibitem[{Pet()}]{Pethick}
\bibinfo{note}{C. J. Pethick and H. Smith, {\it Bose-Einstein Condensation in
  Dilute Gases} (Cambridge University Press, 2008)}.

\bibitem[{Pit()}]{Pitaevski}
\bibinfo{note}{L. Pitaevski and S. Stringari, {\it Bose-Einstein Condensation}
  (Clarendon Press, 2003)}.

\bibitem[{Dut()}]{Dutta}
\bibinfo{note}{A. Dutta {\it et al.}, {\it Quantum Phase Transitions in
  Transverse Field Spin Models} (Cambridge University Press, 2015)}.

\bibitem[{Por()}]{Porras}
\bibinfo{note}{D. Porras and J. I. Cirac, Phys. Rev. Lett. {\bf 92}, 20
  (2004)}.

\bibitem[{You()}]{Young}
\bibinfo{note}{A. P. Young, Phys. Rev. B {\bf 56}, 18 (1997)}.

\bibitem[{Yan()}]{YanHe}
\bibinfo{note}{Y. He and H. Guo, J. Stat. Mech. (2017) 093101}.

\bibitem[{Pla()}]{Platini}
\bibinfo{note}{T. Platini {\it et al.}, J. Phys. A {\bf 40}, 1467 (2007).}

\bibitem[{Pfa()}]{Pfaffian}
\bibinfo{note}{J.-G. Luque and J.-Y. Thibon, Adv. Appl. Math. {\bf 29}, 620
  (2002)}.

\bibitem[{Mat()}]{Mathematica}
\bibinfo{note}{Wolfram Research, Inc. Mathematica version 11.0 (Champain, IL)}.

\bibitem[{nu()}]{nu}
\bibinfo{note}{S.Suzuki {\it et al.}, {\it Quantum Ising Phases and Transitions
  in Transverse Ising Models} (Springer, 2013)}.

\bibitem[{pyt()}]{python}
\bibinfo{note}{E. Jones {\it et al.}, {\it SciPy: Open Source Scientific Tools
  for Python, 2001-}, http://www.scipy.org/ [Online; accessed 01/10/2018]}.

\end{thebibliography}

\end{document}